\documentstyle[12pt,psfig]{l-aa}
\input tcilatex
\newcommand{\beq}{\begin{equation}}
\newcommand{\eeq}{\end{equation}}
\newcommand{\ba}{\begin{array}}
\newcommand{\ea}{\end{array}}

\newcommand{\lapprox}{\lower 3pt\hbox{$\buildrel < \over \sim\;$}}
\newcommand{\gapprox}{\lower 3pt\hbox{$\buildrel > \over \sim\;$}}

\begin{document}

\thesaurus{}    
\title{Flat FRW Cosmologies with Adiabatic Matter Creation: Kinematic tests}
\author{J. A. S. Lima and J. S. Alcaniz}

\institute{Departamento de Fisica, UFRN, C.P 1661\\
	   59072-970 Natal, Brasil }

\date{Received ; accepted}    

\offprints{J. A. S. Lima}      

\maketitle

\begin{abstract} 
Some observational consequences 
of a cosmological scenario driven by adiabatic matter 
creation are investigated. Exact expressions for 
the lookback time, age of the universe, 
luminosity distance, angular diameter,  
and galaxy number counts redshift relations are derived 
and their meaning discussed in detail. The expressions 
of the conventional FRW models are significantly modified 
and provide a powerful method to limit the 
parameters of the models.
\end{abstract}

Nowadays, the increasing difficulties of 
the standard Friedmann-Robertson-Walker
(FRW) cosmologies compel the investigation of   
alternative big-bang models. One of the main
observational motivations 
is the conflict between the expanding age of 
the universe, as inferred from recent measurements of the Hubble 
parameter (Friedmann 1998), and the 
age of the oldest stars in globular 
clusters (Bolton and Hogan 1995, Pont et al. 1998). 
The corresponding uncertainities related with such determinations are now believed to be considerably small, thereby ruling out large regions in the space parameter of the standard cosmology even for open models (Bagla, Padmanabhan and Narlikar 1996). Such restrictions may be even more efficient near future, first, from better data analysis with consequent reduction in incertainities, and second, due to improved experiments as well as new observational facts. For instance, the recent discovery of a $3.5Gyrs$ old galaxy at $z=1.554$ (Dunlop et al. 1996) has been proved to be incompatible with ages estimates for a flat universe unless the Hubble parameter is less than $45kms^{-1}Mpc^{-1}$. Such a constraint is  more stringent than  globular cluster age constraints (Krausss 1997). This ``age problem" is not an isolated difficulty of the FRW model since it also affect others basic features of the standard cosmology, like the structure formation through gravitational amplification of small primeval density perturbations (Jeans' Instability). Indeed, with exception of a very low Hubble constant variant this galaxy formation scenario seems to be  unconsistent with the estimated age of the universe. For open universes, where the ``age problem" is less acute, this happens because the growth of perturbations since recombination is relatively supressed in a low density models (Kofmann, Gnedin and Bahcall 1993, White and Bunn 1995). 

On the other hand, recent measurements of the decceleration parameter using 
Type Ia supernovae (Garnavich et al. 1998, Perlmutter et al. 1998) indicate that the universe may be accelerating today, or equivalently, that the decceleration parameter may be negative. These measurements pose a big problem to the standard model (for any of its variants) since their predictions are $q_{o} > 0$, whatever the sign adopted for the curvature of the model. More recently still, improved observations from a sample of 16 supernovae type Ia plus 34 nearby novae were used to place constraints in $H_o$, $\Omega_m$, $\Omega_{\Lambda}$ and $q_o$ (Riess et al 1998). These authors concluded that the standard flat model ($\Omega_m=1, \Omega_{\Lambda}=0, q_o > 0$) is ruled out and that several effects, among them, extinction , evolution, sample selection bias, local flows are not enough to reconcile the data with the predictions of this model. 
In such state of affairs, it seems more prudent to follow the tradition in cosmology by thinking about alternative big-bang scenarios. As a matter of fact, the positive evidences to standard model, although not negligible are not at all abundant, and are presently under suspection.    
 
Some years ago, a thermodynamic description 
for gravitational creation  of matter and radiation has been proposed in the 
literature (Prigogine et al. 1989; Lima, Calv\~ao 
and Waga 1991; Calv\~{a}o, Lima and Waga 1992; Lima and Germano 1992). 
The crucial ingredient of this formulation 
is the explicit use of a 
balance equation for the number density of the
created particles in addition to Einstein field equations(EFE). In this framework, the thermodynamic second 
law leads naturally to a reinterpretation of  the  energy momentum tensor(EMT) corresponding to an additional 
stress term (creation pressure), which in turn depends on the matter creation rate, and  may alter 
considerably several predictions of the standard 
big-bang cosmology. The issue related to the 
compatibility between this 
approach and the kinetic theory of a relativistic 
gas has also been 
addressed (Triginer, Zimdahl and Pav\'{o}n 1996, Zimdahl, Triginer and Pav\'{o}n 1996).
Studies envolving  matter creation and the early universe physics include the singularity problem 
(Prigogine et al. 1989; Abramo and Lima 1996), reheating 
during the inflationary epoch (Zimdahl and Pav\'on 1994), 
the age of the universe problem (Lima, Germano and Abramo 1996), the entropy problem (Lima and Abramo 1996; Brevik and Stokkan 1996) and the amplification 
of gravitational waves (Maia, Carvalho and Alcaniz 1997; Tavares and Maia 1998). 
Of special interest to us is the particular case termed ``adiabatic'' 
matter creation. Under ``adiabatic'' conditions, particles 
(and consequently entropy) are continuously generated, however, the specific entropy per particle of each component remains constant during the whole process (Calv\~ao, Lima and Waga 1992; 
Lima and Germano 1992). In particular, for photon creation this means that the equilibrium relations are 
preserved ($n \sim T^3$, $\rho \sim T^4$) and also that the photon spectrum may be compatible with the present isotropy of the CMBR (Lima 1996, 1997). 
In addition to the already quoted papers, several authors have also used  such a formulation to study
different dynamical aspects of cosmological models in the last few years 
(Zimdhall and Pav\'on 1993, Brevik and Stokkan 1996). The advantages of this new thermodynamic approach over the old bulk viscosity description 
for matter creation (Zeldovich 1970) have been discussed both in the context of the first order nonequilibrium thermodynamics (Lima and  Germano 1992) and causal formulation (Gariel and Le Denmat 1995). These studies revealed clearly that matter creation cannot consistently be modeled by the bulk viscosity mechanism even considering that both are scalar processes.

An overlook in the literature show, however, that the dynamical properties of cosmological models with ``adiabatic" matter creation have been more carefully investigated than their observational consequences in the present matter dominated phase. This is an important point, because the question for viability of big-bang models with matter creation could partially
be answered deriving expressions for the classical 
cosmological tests, thereby  analysing  the 
influence of this mechanism on the well known predictions of the standard FRW model.  

In order to fill this gap, we foccus our attention here on the quantities of astrophysical interest to the present dust like stage. 
The outline of the paper is as follows. In section 2,
we set up the 
cosmological equations with adiabatic matter 
creation, reviewing briefly some basic 
features of such an approach. 
In section 3, by adopting a creation scenario 
recently proposed (Lima, Germano and Abramo 1996),
we derive new expressions for the observable quantities and analyse some of their properties. The data are then used to limit the unique free quantity (creation parameter) of the model. We conclude with a discussion of the main results.

\section{Flat FRW Equations With ``Adiabatic'' Matter Creation}

Let us now consider the flat FRW line element $(c=1)$
\beq
 ds^2 = dt^2 - R^{2}(t) (dr^2 + r^2 d \theta^2 +
 r^2 sin^{2}\theta d \phi^2) \quad,
\eeq
where $r$, $\theta$, and $\phi$ are dimensionless comoving coordinates and 
$R$ is the scale factor.

 In that background, the nontrivial EFE for a fluid endowed with ``adiabatic"
matter creation and the balance equation for the particle number density can be written as (Prigogine et al. 1989; Calv\~ao, Lima and Waga
1992)
\beq
8\pi G \rho = 3 {\dot{R}^2 \over R^2}  \quad, 
\eeq
\beq
8\pi G (p+p_{c}) = -2 {\ddot{R} \over R} - {\dot{R}^2 \over R^2}  \quad, 
\eeq
\beq
 {\dot{n} \over n} + 3 {\dot{R} \over R} = 
           {\psi \over n} \quad,  
\eeq
where an overdot means time derivative and $\rho$, $p$, $n$ and $\psi$ are the energy density, thermostatic pressure, particle number density and matter creation rate, respectively. The creation pressure 
$p_{c}$ depends on the matter creation rate, and 
for ``adiabatic'' matter creation, it assumes the 
following form (Calv\~ao, Lima and Waga 1992; Lima and Germano 1992)
\beq
p_{c} = - {\rho + p \over 3nH} \psi \quad,
\eeq
where $H = {\dot {R}}/R$ is the Hubble parameter.

As usual in cosmology, the cosmic fluid obeys the ``gamma-law'' equation 
of state 
\beq
 p = (\gamma - 1)\rho \quad,
\eeq   
where the constant $\gamma$ lies on the interval [0,2]. 
 
Combining Eqs. (2) 
and (3) with (5) and (6) it is readily 
seen that the 
scale factor satisfies the generalized FRW differential equation
\beq
R\ddot{R} + ({3 \gamma_* - 2 \over 2}) \dot{R}^2 = 0 \quad,
\eeq
where $\gamma_{*}$ is an effective ``adiabatic index'' 
given by
\beq
\gamma_* = \gamma (1- {\psi \over 3nH}) \quad.
\eeq

To proceed further it is necessary to assume a physically reasonable expression 
to the matter creation rate $\psi$. As can be seen from (4), the dimensionless parameter 
${\psi \over 3nH}$ 
is the ratio between the two relevant rates 
envolved in the process. When this ratio 
is very small 
the creation process can be neglected, and if it is much bigger than unity, we see from (5) that $p_c$ becomes meaningless, because it will be much greather than the energy density. A reasonable upper limit of 
this ratio should be the unity ($\psi=3nH$), since in this case $\psi$ has exactly the value that compensates for the 
dilution of particles due to expansion. In this work we confine 
our attention to the simple phenomenological expression (Lima, Germano and Abramo 1996)  
\beq    
\psi = 3 \beta n H \quad, 
\eeq
where $\beta$ is smaller than unity, and presumably 
given by some 
particular quantum mechanical model for gravitational matter creation. In general, $\beta$ must be a function of the cosmic era, or equivalently, of the $\gamma$ parameter, which specifies  if the universe is vacuum ($\gamma=0$), radiation ($\gamma={4 \over 3}$) or dust ($\gamma=1$) dominated. However, for the sake of brevity we denote all of them generically by $\beta$, assumed here to be a constant at each phase.  

With this choice, the
FRW equation for $R(t)$ given by (7) can be 
rewritten as
\beq
R \ddot{R} + \Delta  \dot{R}^2  = 0 \quad,
\eeq
the first integral of which is
\beq 
{\dot{R}}^2 =  {A \over R^{2\Delta}}  \quad,
\eeq
where $\Delta = {3\gamma(1-\beta)-2 \over 2}$, and from (2) $A$ is a positive constant, 
which must be determined in terms of the present day quantities. It is worth notice that  
for $\beta \geq 1 - {2 \over 3\gamma}$, or equivalently, $\Delta \leq 0$, the above equations imply that $\ddot R \geq 0$, thereby leading to power law inflation. In particular, for $\Delta=0$, these universes expand with constant velocity, and are new examples of coasting cosmologies whose dynamic behavior is driven by matter creation. The observational consequences of ``coasting cosmologies" generated by exotic ``K-matter", like cosmic strings, have been studied in detail (Gott and Rees 1987; Kolb 1989). All of them are characterized by the fact the energy density $\rho \sim R^{-2}$ and the total pressure $P_t= - {1 \over 3}\rho$ (see equations (12) and (15)). 

Using equation (11), it is straightforward to obtain the 
energy density, the pressures ($p$ and 
$p_c$) and the particle number density as 
functions solely of the scale
factor $R$ and of the $\beta$ parameter. These quantities are given by:
\beq
\rho = \rho_o 
{ ( {R_o \over R} ) }^{3\gamma(1-\beta)} \quad,
\eeq
\beq
p_c = - \beta \gamma \rho_o 
{ (  {R_o \over R} ) }^{3\gamma(1-\beta)} \quad,
\eeq
\beq
n=n_o { (  {R_o \over R} )}^{3(1-\beta)} \quad,
\eeq
\beq
P_t = (\gamma_{*} -1) \rho = [\gamma(1-\beta) -1 ] \rho_o 
            {( {R_o \over R} )}^{3\gamma(1-\beta)} \quad,
\eeq
In the above expressions the subscript ``o'' refers to the present values of the parameters, and the total 
pressure is $P_t=p+p_c$. As expected, for $\beta=0$, equations (12)-(15) reduce to those of the standard FRW flat model for all values of the $\gamma$ parameter (Kolb and Turner 1990).

The solution of (11) for all values of 
$\gamma$ and $\beta$
can be written as
\beq
R= R_o { [ 1 + {3\gamma (1-\beta ) \over 2} 
H_o (t-t_{o}) ] }^{{2 \over 3\gamma (1-\beta )}} \quad.
\eeq

Note also that for $\gamma > 0$, we can 
choose $t_{o} =
2 H_{o}^{-1} / 3\gamma (1-\beta )$, with the above 
equation assuming a more familiar form,
namely:
\beq
R(t) = R_o { [ {3\gamma (1-\beta ) \over 2} H_o t ]}^
   {{2 \over 3\gamma (1-\beta )} } \quad.
\eeq
In particular, for a ``coasting cosmology" driven by matter creation one finds $R \sim t$ as it should be. Note also that in the limit $\beta =0$, equations (16) and (17) reduce to the well known
expressions of the FRW flat model.

\section{Expressions for the Observational Quantities}

In what follows we assume that the 
present material 
content of the universe is 
dominated by a pressureless nonrelativistic gas (dust).
Following standard lines we also define
the physical parameters $q_o = - {{R\ddot R \over {\dot R}^2}}\vert_{t=t_o}$ 
(deacceleration parameter) and $H_o={\dot R \over R}\vert_{t=t_o}$ 
(Hubble parameter). From  
(10) it is readily seen that
\beq
q_o = {{1-3\beta}\over 2} \quad.
\eeq
Therefore, for a given value of $\beta$, the decceleration parameter $q_o$ with matter creation is always smaller than the corresponding one of the 
FRW flat model. The critical case ($\beta={1 \over 3}, q_o = 0$), describes a ``coasting cosmology". Curiosly, instead of being supported by ``K-matter" (Kolb 1989), this kind of model is obtained in the present context for a dust filled universe. It is also interesting that even negative values of $q_o$ are allowed for a dust filled universe, since the constraint $q_o < 0$ can always be satisfied provided $\beta > 1/3$. These results are in line with recent measurements of the decceleration parameter $q_o$ using Type Ia supernovae (Perlmutter et al. 1998, Garnavich et al 1998, Riess et al 1998). Such observations indicate that the universe may be accelerating today ($q_o < 0$), which corresponds dynamically to a negative pressure term in the EFE. This would also indicate that the universe is much older than a flat model with the usual decceleration parameter $q_o = 0.5$, and reconcile other recent
results (Freedman 1998), pointing to a Hubble parameter $H_o$ larger than $50$ km s$^{-1}$ Mpc$^{-1}$ (see discussion below Eq.(21)). To date, only models with a cosmological constant, or the so-called ``quintessence" (of which $\Lambda$ is a special case), or still a second dark matter component with repulsive self-interaction have been invoked as being capable of explaining these results (Steinhardt et al. 1997, Cornish and Starkman 1998). In the present context, these prescriptions for alternative cosmologies are replaced by a flat model endowed with an ``adiabatic" matter creation process. Before continuing, we need to express the constant A in terms of  $R_o$ and $H_o$. From (8) one finds
\beq
A=H_o^{2} R_o^{3(1-\beta)} \quad.
\eeq

The kinematical relation distances must be confronted with the observations in order to put limits on the free parameter of the models. 
\\
\\
a) Lookback Time-Redshift

For a given redshift $z$, the scale function $R(t_z)$ is related with $R_o$
by  $1 + z = {R_o \over R}$. The lookback time is exactly the time interval required by the universe to evolving between these two values of the scale factor. Inserting the value of $A$ given above in the first integral (11), the lookback time-redshift relation can be easily derived and it is given by
\beq
t_o - t(z)= {{2H_o^{-1} \over 3(1 - \beta)}\left[1 - {1 \over (1 + z)^{3(1 - \beta) \over 2}}\right]} \quad,
\eeq
which generalizes the standard FRW flat result (Sandage 1988). In figure 1 we have plotted the lookback time as a function of the redshift for some selected values of $\beta$.

Taking the limit $z\rightarrow\infty$ in (20) the present age of the universe (the extrapolated time back to the bang) is  
\beq
t_o = \frac{2H_o^{-1}}{3(1-\beta)} \quad,
\eeq
which reduces for $\beta = 0$ to the same expression of the standard dust model (Kolb and Turner, 1990). 
\begin{figure}
\vspace{.2in}
\centerline{\psfig{figure=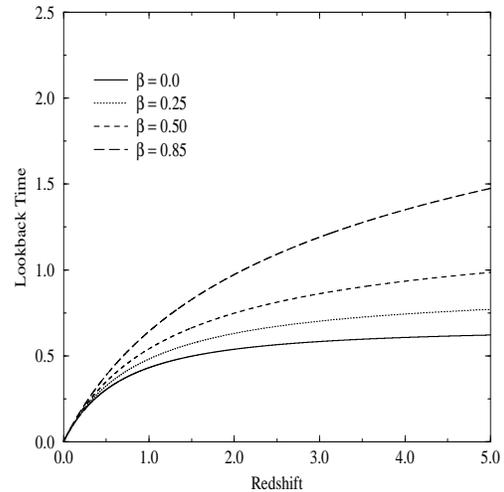,width=3truein,height=3truein}
\hskip 0.1in}
\caption{Lookback time as a function of the redshift for some selected values of $\Omega_{o}$ and $\beta$. Solid curve is the FRW flat universe with no matter creation ($\beta = 0$). The lookback time increases for higher values of $\beta$, i.e., models with larger matter creation rate are older.} 
\end{figure}

Estimates of the Hubble expansion parameter from a variety of methods are now  converging to $h \equiv (H_{o}/100 \rm{km/sec/Mpc})=0.7 \pm 0.1$ (Freedman 1998).  Assuming no matter creation ($\beta = 0$), the lower and upper limits of this value imply that the expansion age of a dust-filled flat universe , which is theoretically favored by inflation, would be either $10.8 \times 10^{9}$ years or
$8.2 \times 10^{9}$ years. These results are in
direct contrast to the measured ages of some stars and stellar systems,
believed to be at least $(12-16)\times 10^{9}$ years 
old or even older if one adds a realistic 
incubation time (Bolte and Hogan 1995, Pont et al. 1998). As can easily be seen from (21), the matter creation process helps because for a given Hubble parameter $H_o$ the expansion age $t_o$ is always larger than ${2 \over 3}H_o^{-1}$, which is the age of the universe for the FRW flat model. It is exactly $H_o^{-1}$ for a coasting cosmology ($\beta = {1 \over 3}$), and greater than $H_o^{-1}$ for $\beta > {1 \over 3}$. In this way, one may conclude that the matter creation ansatz (9) changes the predictions of standard cosmology, thereby alliviating the problem of reconciling observations with the inflationary scenario. It is interesting that matter creation increases the dimensionless parameter $H_{o} t_{o}$ while preserving the overall expanding FRW behavior.
\\
\\
b) Luminosity Distance-Redshift

The luminosity distance of a light source is defined as 
$d_l^{2} = {L \over 4 \pi l}$, where $L$ and $l$ are the absolute and apparent luminosities respectively. In the standard FRW metric (1) it takes the form (Sandage 1961; Weinberg 1972)
\beq
d_l = R_o r(z)(1 + z) \quad,
\eeq
where $r(z)$ is the radial coordinate distance of the object at light emission. Starting from (1), this quantity can be easily derived as follows: since a light signal satisfies the geodesic equation of motion $ds^{2} = 0$ and geodesics intersecting $r_{o} = 0$ are lines of constant $\theta$ and $\phi$, the geodesics equation can be written as
\beq
\int_{o}^{r} dr = \int_{R(t)}^{R_{o}} \frac{dR(t')}{\dot{R}(t')R(t')} \quad.
\eeq
Now, substituting (11) with the value of $A$ given by (19) in the above equation, the radial coordinate distance as function of redshift is given by
\beq
r(z) = \frac{2}{(1-3\beta)R_{o}H_{o}}\left[1-(1+z)^{\frac{3\beta-1}{2}}\right] \quad,
\eeq
and therefore, the luminosity distance-redshift relation is written as
\beq
 H_{o}d_l = \frac{2}{(1-3\beta)}\left[(1+z) - (1+z)^{\frac{1+3\beta}{2}}\right] \quad.
\eeq 
As one may check, taking $\beta = 0$, the above expression reduces to
\beq
H_{o} d_l = 2\left[(1+z) - (1+z)^{\frac{1}{2}}\right] \quad,
\eeq
which is the usual FRW result (Weinberg 1972). On the other hand, expanding (25) for small redshifts after some algebra one finds 
\beq
H_o d_l = z + {1 \over 2}(1 - {{1 - 3\beta} \over 2}) 
z^{2} +... \quad,
\eeq
which depends explicitly on the matter creation $\beta$ parameter. However, inserting (18) we recover the usual FRW expansion for small redshifts, which depends only on the effective deacceleration parameter $q_o$ (Weinberg 1972; Kolb and Turner 1990). The luminosity distance as a function of the redshift is shown in figure $2$. As expected, in these diagrams different models has the same behavior at $z << 1$ (Hubble law), and the possible discrimination among different models comes from observations at large redshifts ($z \geq 1$). However, it is usually believed that in such scales evolutionary effects can not be neglected. The range of possible data at the limiting $z$, for which evolutionary effects are not important are indicated by the data point and error bar (Kristian, Sandage and Westphal 1978).
\begin{figure}
\vspace{.2in}
\centerline{\psfig{figure=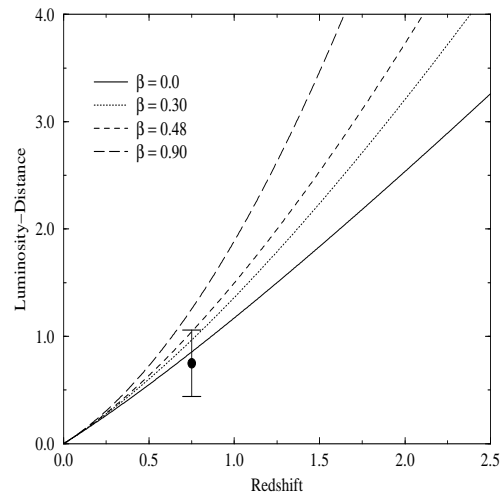,width=3truein,height=3truein}
\hskip 0.1in}
\caption{Luminosity distance as a function of the redshift for flat models with adiabatic matter creation. Solid curve is the Einstein-de Sitter model ($\beta = 0$). The selected values of $\beta$ are shown in the picture. Here the typical error bar and data point are taken from Kristian et al.} 
\end{figure}
\\
\\
c) Angular Diameter-Redshift

The angular size $\theta$ of an object is an extremely sensitive function of 
the cosmic dynamics. In particular, the apparent continuity of the $\theta(z)$ relation for galaxies and quasars is also believed to be a strong support to the cosmological nature of the redshifts (Kapahi 1987). Here we are interested in angular diameters of light sources described as rigid rods and not isophotal diameters. These quantities are naturally different, because in an expanding world the surface brightness varies with the distance (for more details see Sandage 1988).

Let $D$ be the intrinsic size of a source located at $r(z)$, assumed independent of the redshift and perpendicular to the line of sight. If it emits photons at time $t_1$ that at time $t_o$ arrive to an observer located at $r=0$, its angular size at the moment of reception is defined by (Sandage 1961)
\beq
\theta = {D(1 + z) \over R_o r(z)} \quad.
\eeq
Inserting  the expression (24) for $r(z)$ into (28) we find 
\beq
\theta = {DH_o(1-3\beta)(1 + z)^{3(1 -\beta) \over 2} \over {2\left[(1 + z)^{1 - 3\beta \over 2} - 1\right]}} \quad.
\eeq
 A log-log plot of angular size versus redshift is shown in figure $3$ for selected values of $\beta$.

For all models, the angular size initially decreases with increasing $z$, reaches its minimum value at a given $z_c$, and eventually begins to increase for fainter magnituds. This generic behavior for an expanding universe was predicted long ago in the context of the standard model (Hoyle 1959). It may qualitatively be understood in terms of an expanding space: the light observed today from a source at high $z$ was emitted when the object was closer. How this effect depends on the $\beta$ parameter? As can be seen  from (29) the minimal value of which occur at $z_c(\beta) = [{3(1 - \beta) \over 2}]^{2 \over 1 - 3\beta} -1$. Hence, the minimum persists in the presence of adiabatic matter creation, and is pushed to the right direction, that is, it is displaced to higher redshifts as the $\beta$ parameter is increased. As expected, for $\beta = 0$ one finds $z_c = {5 \over 4}$, which is the standard result for a dust filled FRW flat universe. It is also convenient to consider the limit of small redshifts in order to clarify the role played by $\beta$. Expanding (29) we have
 $z$ \beq
\theta = {D H_o \over z}\left[ 1 + {1 \over 2}
(3 + {{1 - 3\beta} \over 2}) z +...\right] \quad.
\eeq
Hence, ``adiabatic" matter creation as modelled here also requires an angular size decreasing as the inverse of the redshift for small $z$. However, the second order term is a function of the $\beta$ parameter. Its overall effect on the angular size is depart it from the Euclidean behavior ($\theta \approx z^{-1}$) more slowly than in the corresponding FRW model (see fig.3). In terms of $q_o$, inserting (18) into (30) it is readily obtained 
\beq
\theta = {D H_o \over z}\left[ 1 + {1 \over 2}
(3 + q_o) z +...\right] \quad,
\eeq 
which is formally the same FRW result for small redshifts (Sandage 1988).  Note that even at this limit, constraints on the decceleration parameter  from the data are equivalent to place limits on the values of 
$\beta$ (see (18)).
\\
\\  
d) Number Counts

Let us now derive the galaxy number per redshift interval in the presence of adiabatic matter creation. We first remark that although modifying the evolution equation driving the amplification of small perturbations, and so the usual adiabatic treatment for galaxy formation, the created matter is smeared out and does not change the total number of sources present in the nonlinear regime. In other words, the number of galaxies already formed scales with $R^{-3}$.

Let $n_g(z,L)dL$ be the proper concentration of sources at redshift $z$  with absolute luminosity between $L$ and $L + dL$. The total number of galaxies $N_g(z)$ is proportional to the the comoving volume
\beq
dN_{g}(z) = n_{g} dL dV_{c} = 4\pi n_{g} r^{2} dr dL \quad.
\eeq

Now, by using that ${dt \over R(t)} = {dR \over R\dot R} = - dr$, we find that
\beq
{dN_{g} \over 4\pi n_g dzdL} = {(R_oH_o)^{-1}r(z)^{2}  
\over \left[(1 + z)^{3(1 - \beta)/2}\right]}  \quad,
\eeq
where $n_g(z,L)=n_o(L)(1 + z)^{3}$.

On the other hand, since the radial coordinate r(z) is given by eq.(24) it follows that the expression for number-counts can be written as
\beq
{(H_oR_o)^{3}dN_{g}\over 4\pi n_oz^{2}dzdL} = \frac{\delta^{2}\left[1-(1+z)^{-\frac{(1-3\beta)}{2}}\right]^{2}}{z^{2}(1+z)^{\frac{3(1-\beta)}{2}}} \quad,
\eeq
where $\delta = \frac{2}{(1-3\beta)}$.
For small redshifts, we have that
\beq
{(H_oR_o)^{3} dN_{g} \over 4\pi n_oz^{2}dzdL} = 
1 - 2\left[{(1-3\beta) \over 2} + 1\right]z + ...  \quad.
\eeq
The number count-redshift diagram for a dust-filled model with ``adiabatic" matter creation is shown in the figure $4$,  for the indicated values of $\beta$. Table 1 summarizes the limits to $\beta$ obtained from each kinematical test.
\begin{figure}
\vspace{.2in}
\centerline{\psfig{figure=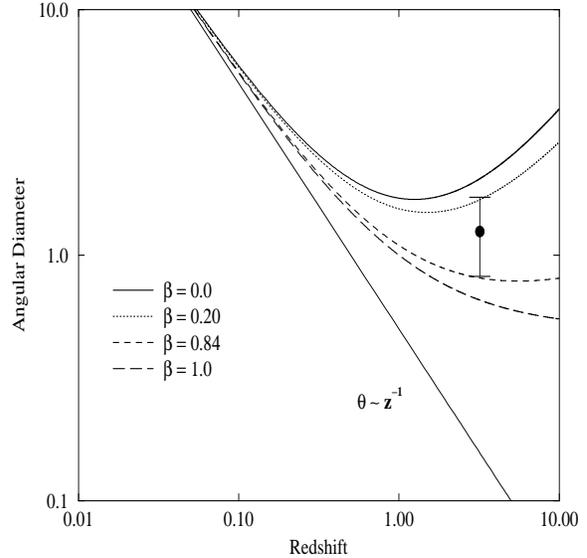,width=3truein,height=3truein}
\hskip 0.1in}
\caption{Angular diameter versus redshift in flat models with adiabatic matter creation and somo selected values of $\beta$. Solid curve is the standard model ($\beta = 0$). The angular size reaches a minimum at a given $z_{c}$ and increases for fainter magnitudes. The minimum is displaced for higher $z$ as the $\beta$ parameter is increased. The typical error bar and data point are taken from Gurvits (1994).} 
\end{figure}  
\begin{figure}
\vspace{.2in}
\centerline{\psfig{figure=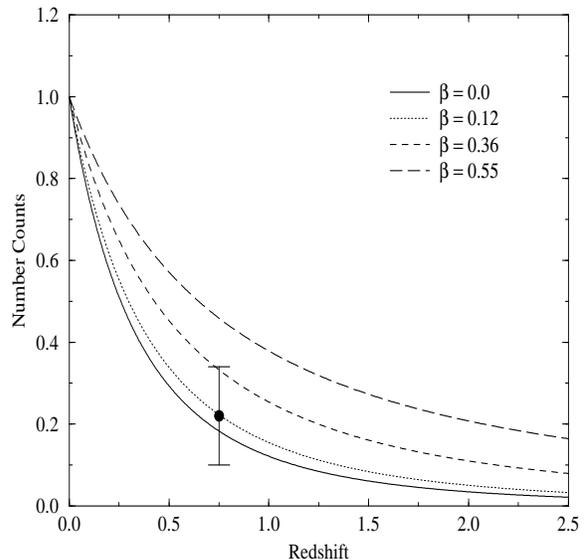,width=3truein,height=3truein}
\hskip 0.1in}
\caption{Number counts as a function of the redshift for flat models with adiabatic matter creation. Solid curve is the standard Einstein-de Sitter model. The selected values of $\beta$ are shown in the picture. Typical error bar and data point are taken from Loh and Spillar (1986).} 
\end{figure}

\section{Conclusion}

The cosmological principle (homogeneity and isotropy of space) defines the shape of the line element up to a spatial scale function, which must be time dependent from the cosmological nature of the redshifts. As discussed here, the expanding ``postulate'' and its main consequences may also be compatibilized with a cosmic fluid endowed with adiabatic matter creation. The similarities and differences among universe models with matter creation as described in the new thermodynamic approach and the conventional matter conserved FRW model have been analysed both from formal and observational view points. The rather slight changes introduced by the matter creation process, which is quantified by the $\beta$ parameter, provides a reasonable fit of some cosmological data. Kinematic tests like luminosity distance, angular diameter and number-counts versus redshift relations constrain perceptively the matter creation parameter (see table 1). For flat models with $\beta \neq 0$, the age of the universe is always greather than the corresponding FRW model ($\beta=0$). More important still, the decceleration parameter $q_o$ may be negative as suggested by recent type Ia supernovae observations. In this concern, the models studied here are alternatives to universes dominated by a cosmological constant or ``quintessence". 

The angular size versus redshift curves have the minimum displaced for higher values of $z$, thereby alliviating the problem in reconciling the angular size data from galaxies and quasars at intermediate and large redshifts. It is also interesting that all the theoretical and 
observational results previously obtained whitin a scenario driven by $K$-matter (Kolb 1989), are reproduced for a dust-filled universe with $\beta = {1 \over 3}$. 

We also stress that in spite of these important physical consequences, the present day matter creation rate, $\psi_o = 3n_oH_o \approx 10^{-16}$ nucleons $cm^{-3}yr^{-1}$, is nearly the same rate predicted by the steady-state universe (Hoyle, Burbidge and Narlikar 1993).  This matter creation rate is presently far below detectable limits.

The constraints on the $\beta$ parameter should be compared with the corresponding ones using the predictions of light elements abundances from primordial  nucleosynthesis. In fact, the important observational quantity for nucleosynthesis is the baryon to entropy ratio. In these models the temperature scale-factor relationship and entropy density are modified, therefore one may expect sensitive implications to the nucleosynthesis scenario. 
 
Finally, we remark that it is not so difficult to widen the scope of the kinematic results derived here to include curvature effects as well as a non-zero cosmological constant. In particular, concerning the ``age problem", even closed universes seems to be compatible with the ages of the oldest globular clusters, when the value of the creation parameter is suficiently high. Further details about kinematic tests in closed and opened universes with matter creation  will be published elsewhere (Alcaniz and Lima 1999).
\begin{table}[t]
\caption{Limits to $\beta$}
\begin{tabular}{rl} 
\hline
\\
\multicolumn{1}{c}{Test}&
\multicolumn{1}{c}{$\beta$}\\
\\
\hline
\\
Luminosity distance-redshift& 
$\beta \le 0.48$\\
\\
Angular size-redshift&  
$0.20 \le \beta \le 0.84$\\
\\
Number counts-redshift&  
$\beta \le 0.36$\\
\\
\hline
\end{tabular}
\end{table}

\begin{acknowledgements}
One of the authors (JASL) is grateful to Raul Abramo for helpful discussions. 
This work was partially supported by the project Pronex/FINEP (No. 41.96.0908.00) and Conselho Nacional de Desenvolvimento Cient\'{\i}fico e
Tecnol\'ogico - CNPq (Brazilian Research Agency).
\end{acknowledgements}


\end{document}